\documentclass[letterpaper,twocolumn,aps,prb,groupedaddress,superscriptaddress,showpacs,amsfonts]{revtex4}

\usepackage{epsfig}
\usepackage{graphicx}
\usepackage{dcolumn}
\usepackage{bm}

\DeclareMathAlphabet{\mathpzc}{OT1}{pzc}{m}{it}


\newcommand{\beq}{\begin{equation}}
\newcommand{\eeq}{\end{equation}}
\newcommand{\bea}{\begin{eqnarray}}
\newcommand{\eea}{\end{eqnarray}}

\begin{document}

\title{On the Validity of the Tomonaga Luttinger Liquid Relations for the One-dimensional Holstein Model}

\author{Ka-Ming Tam}
\altaffiliation{Present address: Department of Physics and Astronomy, Louisiana State University, Baton Rouge, LA 70803}
\affiliation{Department of Physics, Boston University, 590 Commonwealth Ave., Boston, MA 02215}

\author{S.-W. Tsai}
\affiliation{Department of Physics and Astronomy, University of California, Riverside, CA 92521}

\author{D.~K.~Campbell}
\affiliation{Department of Physics, Boston University, 590 Commonwealth Ave., Boston, MA 02215}

\date{\today}

\begin{abstract}

For the one-dimensional Holstein model, we show that the relations among the scaling exponents of various correlation functions of the Tomonaga Luttinger liquid (LL), while valid in the 
thermodynamic limit, are significantly modified by finite size corrections. We obtain analytical expressions for these corrections and find that they decrease very slowly with increasing system 
size. The interpretation of numerical data on finite size lattices in terms of LL theory must therefore take these corrections into account. As an important example, we re-examine the proposed metallic phase of the zero-temperature, half-filled one-dimensional Holstein model {\it without} employing the LL relations. In particular, using quantum Monte Carlo calculations, we study the competition between the singlet pairing and charge ordering. Our results do not support the existence of a dominant singlet pairing state. 

\end{abstract}
\maketitle

\section{Introduction}
The study of quantum fluctuations arising from dynamical phonons and their effects on the charge ordering instability of one-dimensional systems has a long history.\cite{Peierls} One of the simplest 
models to describe a one-dimensional solid under the influence of the quantum fluctuations is the molecular-crystal or Holstein model,\cite{Holstein1,Holstein2} which is given by the following 
Hamiltonian: 
\begin{eqnarray}
H = H_{el} + H_{ph} + H_{el-ph},
\label{hhm}
\end{eqnarray}
where
\begin{eqnarray}
H_{el} &=& - t \sum_{i,s}(c_{i,s}^{\dagger}c_{i+1,s}+ H.c.)  \ ,
\\
H_{ph} &=& \sum_{i}(\frac{p_{i}^{2}}{2M} + \frac{K}{2} q_{i}^{2}) \  ,
\\
H_{el-ph} &=& g_{\rm{ep}} \sum_{i,s}q_{i}n_{i,s}     \ .
\label{hhmterms}
\end{eqnarray}

In this model, the electrons move in a tight-binding band 
interacting with the lattice via the coupling between the local electron density and the local lattice displacement. In the Hamiltonian, 
$c_{i,s}^{\dagger}$ ($c_{i,s}$) creates (annihilates) an electron at site $i$ with spin $s$, and $n_{i,s}$ is the electron number operator. The ions are described by momentum $p_{i}$ and 
displacement $q_{i}$ for site $i$. The mass of the ions is $M$, and $K$ is the
harmonic lattice coupling constant, so that the phonon frequency, $\omega_0 = \sqrt{K/M}$. 
Following convention, we shall call these dispersionless optical excitations ``Einstein phonons".
The hopping energy of the electrons 
between nearest neighbors is $t$, and $g_{\rm{ep}}$ is the electron-phonon coupling energy. In what follows we will use units such that $t=M=1$, except where otherwise specified. For the half-filled band, early work 
suggested that the 
Holstein model displays dimerization and that long range charge ordering sets in at any finite phonon frequency.\cite{Hirsch,Hirsch2,Bourbonnais1,Caron,Schmeltzer,Yonemitsu} These conclusions have 
been challenged by 
various studies which find that for small enough electron-phonon coupling, the system has zero charge 
gap and thus no long range order. \cite{Wu,Jeckelmann,Takada,Takada2,Tezuka,Tezuka2,Tezuka3,Feshke,Ejima,Chatterjee,Feshke2,Hotta,Feshke3,Zhao,H-Zhao}

A recent quantum Monte Carlo study\cite{Clay1} further suggested that the dominant power-law ordering is the singlet BCS superconducting pairing (SS) correlation. This conclusion was based on data from finite size 
simulations which shows that the Luttinger 
charge exponent ($K_{\rho}$) is larger than one.\cite{Schulz} (It is important to realize at the outset that this is not an off-diagonal long range ordering\cite{Yang}, as 
breaking 
the $U(1)$ gauge invariance\cite{Anderson1,Anderson2} to generate a mass gap is forbidden at one dimension.\cite{Mermin-Wagner,Hohenberg,Coleman} SS should thus be understood as a charge
gapless, dominant off-diagonal power-law decaying ordering.) 
In the (possibly singular) limit of infinite phonon frequency, the Holstein model becomes equivalent to the attractive Hubbard model. \cite{Hirsch} For the attractive Hubbard model at half-filling, the SS and the charge density wave (CDW) order parameters are 
degenerate because of the 
particle-hole symmetry. Moreover, both order parameters possess continuous symmetry with respect to the spin rotation.\cite{Hirsch} Therefore, long range ordering of either sort is forbidden in 
this case. 
However, from the numerical simulations for finite size lattices, 
$K_{\rho}$ is always larger than one and converges to one very slowly with increasing system 
size.\cite{Clay2} The similar behaviors of $K_{\rho}$ for both the attractive Hubbard model and the Holstein model at finite phonon frequency have led to the suggestion that the Holstein model 
{\it even 
at finite phonon frequency} may have a charge gapless metallic phase with degenerate SS and CDW correlations, even though $K_{\rho}>1$ from finite size calculations.\cite{Clay2} We note that this 
coexistence has been proposed years ago.\cite{Guinea}

In the present article we present analytical and numerical calculations 
of the LL relations in the 1D Holstein model 
with finite size corrections. 
Our results extend to Einstein phonons 
previous studies involving acoustic phonons\cite{Loss, Martin} and complement prior multiscale functional renormalization group studies of retardation effects due to Einstein phonons.\cite{Tam} 
In essence, we seek to answer three important 
questions raised by previous studies:(I) Are the LL scaling relations valid in the 1D Holstein model?  (II) What are the dominant correlations in this model?  (III) What can be said about the possibility of a charge gapless metallic phase at half-filling for weak electron-phonon coupling? As we shall see, we are able to provide clear answers to the first two questions but can only comment on the third.

In section II of the paper, we begin our study by describing the forward scatterings of the Holstein model in terms of bosonic fields. In section III, we solve analytically the forward scattering 
part of the Holstein model to show that electron-phonon coupling modifies the relations between scaling exponents
, and we calculate these modifications for finite size systems.  This enables us to answer question (I). In section IV, we employ 
quantum Monte Carlo calculations for the full Hamiltonian (Eq. \ref{hhm}) to resolve Question (II) and to gain some insight into Question (III). In section V, we discuss the reasons of why all the perturbative renormalization group studies obtain long range CDW phase and under what circumstances the metallic phase can be obtained. Section VI contains a brief conclusion.

\section{Forward scatterings in the 1D Holstein model}

To answer the question about the validity of the LL relations for the Holstein model in the {\it gapless} regime, {\it including the gapless half-filled case if present}, it is sufficient to 
diagonalize exactly the {\it forward scatterings (scatterings with zero momentum transfer, often denoted as $g_{2}$ and $g_{4}$ in the ``g-ology" terminology)}\cite{Solyom} of the Hamiltonian. 
The neglect of anything beyond forward scatterings is unquestionably an assumption which cannot be {\it a priori} justified in the full model, but this neglect is the fundamental conjecture on the description of the gapless one-dimensional system: any contributions beyond forward scatterings can be incorporated into the renormalization of the forward scatterings. Therefore, although we are considering forward scatterings only, it is completely general for the description of any gapless regime {\it regardless of the filling}.
In what follows, the right hand side of the symbol, $\approx$, is understood to include only those terms in which the forward scatterings {\it exclusively} are retained. 

By formulating the one-dimensional problem in the continuum limit and retaining only the low-energy degrees of freedom, $H_{el}$, can be expressed in term of bosonic fields, $\varphi_{\nu}$ and their 
corresponding momenta  $\Pi_{\nu}$ with explicit spin$(\sigma)$ - charge$(\rho)$ separation.\cite{Emery,Voit}
\begin{eqnarray}
H_{el} &=& H_{\rho} + H_{\sigma},
\\
\nonumber
H_{\nu} &=& \int dx [u_{\nu}K_{\nu}\Pi_{\nu}^{2} + \frac{
u_{\nu}}{K_{\nu}} (\partial_{x}\varphi_{\nu})^{2}].
\label{LLH}
\end{eqnarray}
For the Holstein model at half filling, the Fermi points are at $k_{F}=-\pi/2$ and $\pi/2$. Therefore $u_{\sigma} = u_{\rho} = 2t$, and $K_{\sigma} = K_{\rho} = 1$.\cite{Voit,Loss,Martin} The values 
of these parameters, $u_{\sigma},u_{\rho},K_{\sigma},K_{\rho}$, can be renormalized by the interactions. However, as long as the LL description is valid, the scalings of the various correlation 
functions are determined solely by these parameters.\cite{Voit,Emery} In particular, the CDW correlation and the SS correlation scale as $O_{CDW}(x) \propto 
x^{-K_{\rho}-K_{\sigma}} \equiv x^{\alpha_{CDW}}$ and $O_{SS}(x) \propto x^{-K_{\rho}^{-1}-K_{\sigma}} \equiv x^{\alpha_{SS}}$, respectively. This implies $\alpha_{CDW} \alpha_{SS} = 1$, when a spin 
gap appears, in which case $K_{\sigma}=0$.

Consider the electron-phonon coupling. As we include only forward scatterings, only the phonons with momentum $k=0$ are coupled with the electrons. The Hamiltonian for the phonons with momentum $k=0$ is
\begin{eqnarray}
H_{ph} \approx \frac{1}{2}\int dx [\zeta^{-1}\Pi_{d_{0}}^{2} + \zeta \omega_0^2 d_{0}^{2}], 
\label{LLH_ph}
\end{eqnarray}
where $\zeta$ is the mass density, $d_{0}$ is the phonon field with momentum $k=0$,
and $\Pi_{d}$ is the conjugate momentum. 
The Hamiltonian for the electron-phonon coupling is
\begin{eqnarray}
H_{el-ph} &\approx& g_{\rm{ep}} \sum_{s} \int dx (\psi_{s}^{\dagger} \psi_{s} d_{0}).
\end{eqnarray}
By keeping only the term which is 
linear in the bosonic field to express the density operator, $\psi_{s}^{\dagger}\psi_{s}$, we obtain\cite{Voit,Loss,Martin}
\begin{eqnarray}
H_{el-ph} &\approx& g_{\rm{ep}} \int dx \sqrt{\frac{2}{\pi}}(\partial_{x} \varphi_{\rho}) d_{0}.
\label{LLH_g}
\end{eqnarray} 
The remaining terms will generate the back scatterings and Umklapp scatterings. They also lead to spin-charge coupling due to the retardation.\cite{Voit}

When only forward scatterings are considered, the electron-phonon coupling affects only the charge part of the Hamiltonian, and we will thus focus on the charge part only in the following. The spin 
part is unchanged from the standard Luttinger model.\cite{Voit,Loss,Martin} The charge part of the Hamiltonian (Eq. \ref{hhm}) 
arising solely from forward scatterings is
\begin{eqnarray}
H_{forward} \approx
 \int dx [u_{\rho}K_{\rho}\Pi_{\rho}^{2} + \frac{
u_{\rho}}{K_{\rho}} (\partial_{x}\varphi_{\rho})^{2}] + 
\\ \nonumber
\frac{1}{2}\int dx [\zeta^{-1}\Pi_{d_{0}}^{2} + \zeta \omega_0^2 d_{0}^{2}] +
g_{\rm{ep}} \int dx \sqrt{\frac{2}{\pi}}(\partial_{x} \varphi_{\rho}) d_{0}.
\label{H_forward}
\end{eqnarray} 
Since the phonon fields appear at quadratic order, we follow the work by Martin and Loss,\cite{Loss,Martin} who considered the case of acoustic phonons (as opposed to the Einstein phonon case we are 
considering here) to integrate out the phonon fields, 
$d_{0}$, which results in the collective propagator for the charge field, $\varphi_{\rho}$, 
\begin{eqnarray}
D_{\rho}(k,\omega)= \frac{1}{K_{\rho}u_{\rho}}\left[(\omega^{2}+u_{\rho}^{2}k^2) - \frac{b( k\omega_{0}u_{\rho})^{2}}{\omega^{2}+\omega_{0}^{2}}\right],
\label{Drho}
\end{eqnarray} where $b= 2 K_{\rho} g_{\rm{ep}}^{2}/\pi u_{\rho} \omega_{0}^{2}$. The first part of the propagator is the contribution from the bosonized electrons, and the second part comes 
from the electron-phonon coupling. The crucial difference between Einstein phonons and acoustic phonons is that the coupling between the two bosonic modes ($\varphi_{\rho}$ and $d_{0}$) is symmetric for acoustic phonons ($\partial_{x} \varphi_{\rho} \partial_{x} d_{0}$)\cite{Martin,Loss}, but 
asymmetric for Einstein phonons ($\partial_{x} \varphi_{\rho} d_{0}$). As a result, the dispersion relations for the collective modes are linear for acoustic phonons but nonlinear for Einstein phonons. Therefore, the acoustic phonon case 
can be solved exactly, but the Einstein phonon case, in general, cannot be solved exactly.

\section{Modifications of Scaling Exponents}
It is generally believed that the spin part of the half-filled Holstein model is always gapped due to the back-scatterings.\cite{Hirsch,Clay1,Clay2} (The spin-gapped but charge-gapless 
LL is often referred as Luther-Emery liquid.\cite{Luther-Emery})
As a consequence, the Luttinger spin exponent $K_{\sigma}$  
renormalizes to zero. We take this as an assumption and will not elaborate on it here.
Therefore, the boundary of the transition between the charge ordering and the singlet pairing is determined by the points where the Luttinger charge exponent crosses the value {\it one}, providing that the Luttinger liquid description survives. 

It has previously been shown that the LL scaling relations are modified by the coupling with acoustic phonons.\cite{Loss,Martin} The analogous exact relations cannot be obtained for the Holstein 
model, which has Einstein phonons. Nevertheless, we can still extract the power-law decaying terms {\it asymptotically (i.e., long-distance, low-energy limit, $x \rightarrow \infty, k \rightarrow 
0$)}. We find 
\begin{eqnarray}
  \alpha_{CDW} &=& - \frac{u_{\rho}}{v_{-}} K_{\rho};\\
  \alpha_{SS} &=& - \frac{u_{\rho}}{v_{-} K_{\rho}}(1-b),
\label{K_asy}
\end{eqnarray} to the lowest order in the momentum. At this order, the velocity of the lower branch, $v_{-}$ is $u_{\rho}\sqrt{1-b}$. Hence the relation $\alpha_{CDW} \alpha_{SS} = 1$ is preserved. 
\cite{Chen} We reiterate that these asymptotic results are strictly correct only at long-distance, even apart from any oscillatory contributions. In addition, the power-law decay of the 
correlations is also valid at long distances only.

The exact results for the charge density, $O_{CDW}$, and the singlet pairing, $O_{SS}$, correlation functions can be expressed in term of two series:  
\begin{eqnarray}
  \ln(O_{CDW}(b,x)) &=& \frac{-1}{2\pi}\sum_{k} \frac{uK_{\rho}}{\omega_{-}} 
 \left(1-\frac{\omega_{0}^{2} -\omega_{+}^{2}}{\omega_{-}^{2}-\omega_{+}^{2}}\right)
  \nonumber\\
&& [1-\cos(kx)\exp(-\omega_{-}t)], 
  \\
  \ln(O_{SS}(b,x)) &=& \frac{-1}{2\pi}\sum_{k}\frac{u}{K_{\rho}\omega_{-}}
\left[1-\frac{(1-b)\omega_{0}^{2} -\omega_{+}^{2}}{\omega_{-}^{2}-\omega_{+}^{2}}  \right]
\nonumber\\
&&[1-\cos(kx)\exp(-\omega_{-}t)], 
\label{K_exact}
\end{eqnarray}
where 
\begin{eqnarray}
\omega_{\pm}^{2}(k)\!=\!\frac{\omega_{0}^{2} + (u_{\rho}k)^{2} \pm 
\sqrt{[\omega_{0}^{2}-(u_{\rho}k)^{2}]^{2}+4b(\omega_{0}u_{\rho}k)^{2}}}{2} .
\end{eqnarray}
We have omitted oscillatory contributions. In addition, the upper branch of the collective modes is ignored due to the 
gap with magnitude, $\omega_0$; this differs from the acoustic case \cite{Loss,Martin} in which both branches are gapless and must be included. 

\begin{figure}[bth]
\centerline{
\includegraphics*[height=0.26\textheight,width=0.48\textwidth, viewport=0 0 620 450,clip]{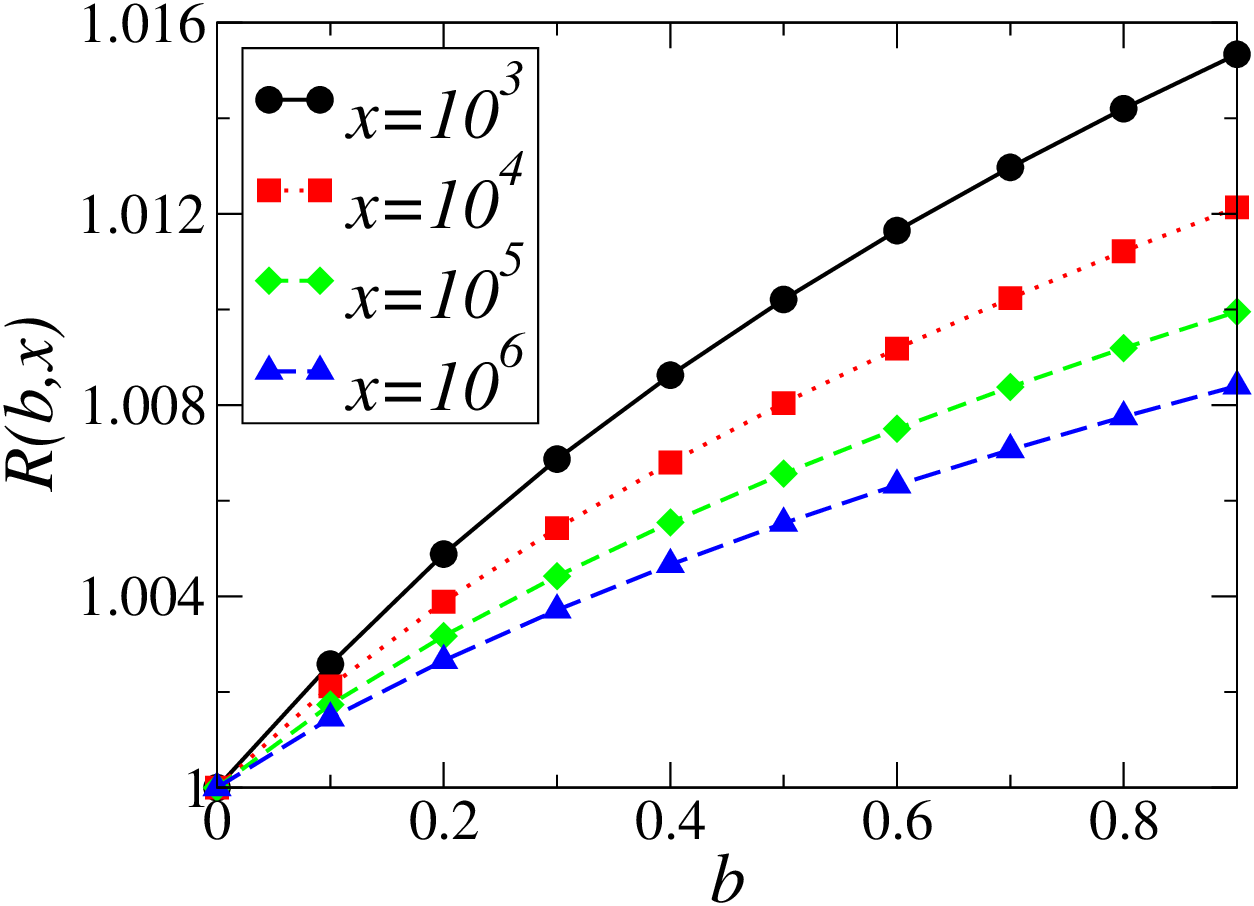}} 
\caption{Plot of the ratio, $R(b,x) \equiv \frac{ln[O_{CDW}(b,x)] ln[O_{SS}(b,x)]}{ln[O_{CDW}(0,x)] ln[O_{SS}(0,x)]}$, versus the scaled electron-phonon coupling, $b$, at various distances, 
$x$, for $t \rightarrow 0^{+}$ }  
\label{LL_exp}
\end{figure}

To investigate the validity of the relation $\alpha_{CDW} \alpha_{SS} = 1$ at finite distance we take the continuum limit, which amounts to replacing the summation over the momentum by an integral; 
and we have to introduce the low momentum cutoff $\epsilon$ \cite{Loss,Martin} and the high momentum cutoff is set to 
$2 u_{\rho}$. 
For the following numerical calculations, we fix the phonon frequency, $\omega_0=1$.  The asymptotic results indicate that the relation, $\alpha_{CDW} 
\alpha_{SS} = 1$, is valid independent of any parameter. In particular, it is valid independent of the electron-phonon coupling, which enters via the
parameter $b$.
Therefore, the ratio of the product of the singlet pairing and the charge ordering correlations at $b\neq0$ and $b=0$,  
\begin{eqnarray}
R(b,x) \equiv \frac{ln[O_{CDW}(b,x)] ln[O_{SS}(b,x)]}{ln[O_{CDW}(0,x)] ln[O_{SS}(0,x)]},
\end{eqnarray}
should be equal to one when $x \rightarrow \infty$, that is, 
\begin{eqnarray}
\lim_{x \rightarrow \infty} R(b,x) =
\frac{\alpha_{CDW}(b\neq0) \alpha_{SS}(b\neq0)}{\alpha_{CDW}(b=0) \alpha_{SS}(b=0)}=1. 
\label{exp_ratio}
\end{eqnarray} 
While Eq. \ref{exp_ratio} is exact in the asymptotic limit, using the analytical expressions 
we obtained for $O_{SS}(b,x)$ and $O_{CDW}(b,x)$, 
we find that 
$R(b,x)$ 
deviates from one for non-zero scaled electron-phonon couplings, $b$; even at rather long distances, $x$. 
This is shown in detail in Fig. \ref{LL_exp}. Importantly, this deviation depends not only on the distance but also on the value of the electron-phonon coupling. 
The ratio is 
clearly larger than one and seems to converge to the asymptotic value, {\it one}, very slowly and at distances that are likely beyond the reach of existing numerical calculations, which typically study systems up to a few hundred sites.
This behavior is also likely to preclude a reliable scaling estimate of its asymptotic value.
Our results suggest that the product of the exponents of the singlet pairing and of the charge ordering is larger 
than one at finite distance for non-zero electron-phonon coupling. Consequently, observing a charge exponent larger than one does not automatically imply that the singlet pairing correlation is the 
dominant instability. {\it Thus although  $\alpha_{CDW} \alpha_{SS} = 1$ in the asymptotic limit, in a finite size system the value of $\alpha_{CDW} \alpha_{SS}$ not only depends on the system size 
but also on the forward scatterings strengths}.

The deviation from one is enhanced by the scaled electron-phonon coupling, $b$, and suppressed by the distance, $x$. These results are consistent with the charge exponent obtained in a Quantum Monte 
Carlo study,\cite{Clay1} in which the charge exponent is larger than one and increases with electron-phonon coupling. In addition, the Quantum Monte Carlo calculations also found that the charge 
exponents decrease with increasing system size.\cite{Clay2}

Our results show that even if all the non-linear operators, which are always present and especially important in the half-filling case, are irrelevant and renormalized to zero and the Holstein model 
is in the gapless regime which can be described in terms of LL, the charge exponent alone {\it at finite distance} cannot resolve the competition between the charge ordering and the singlet superconducting
pairing. In fact, the ``two-cutoff" renormalization group\cite{Grest,Caron,Bourbonnais2,Zimanyi} calculations have shown that the charge gap is always opened by the Umklapp scatterings.\cite{Bindloss} 
Similarly, multiscale renormalization group approaches which systematically renormalize the frequency dependences of the couplings\cite{Tsai,Tsai2,Tsai3} also ruled out the existence of dominant 
singlet 
pairing due to the contribution from the dynamical Umklapp scatterings at high energy.\cite{Tam} Further, a recent study once again confirmed the CDW phase and further discussed the 
quantum-classical crossover.\cite{Bakrim}

From the results of the correlation functions for the forward scatterings of Eq. \ref{hhm}, it is quite clear that using the relation, $\alpha_{CDW} \alpha_{SS} = 1$, to decide the dominating 
correlations in the 
gapless phase 
of the half-filled Holstein model can lead to misleading results. We reiterate that the violation of this relation at finite system sizes is not because of the oscillating terms from normal 
ordering but because of the non-linear collective modes from the asymmetric coupling between the electrons and the dynamical phonons.

\section{Monte Carlo calculations of the correlation functions}

Our previous results make clear that we cannot rely on the LL relations for the thermodynamic limit to determine the correlation exponents from 
finite size simulations of the Holstein model. Hence we turn instead to a numerical study of the full model (Eq. 1) and use the  
projector determinantal Monte Carlo\cite{Blankenbecler,White,Berger,Linden} to determine the ``ground state" and its correlation functions of finite size systems. The ``ground state" ($|\rm{GS}\rangle$) is obtained by projecting the free fermion trial state ($|\Psi_{\rm T}\rangle$) as $|\rm{GS}\rangle = lim_{\beta \rightarrow \infty} exp(-\beta 
{\it H}) |\Psi_{\rm T}\rangle$, in practice $\beta$ is set between $10-20$. We use the $|\rm{GS}\rangle$ to calculate the pairing correlation, $O_{SS}$, and the charge ordering correlation, $O_{CDW}$, in 
the real space directly,

\begin{figure}[tbp]
\centerline{
\includegraphics*[height=0.15\textheight,width=0.23\textwidth, viewport=-5 0 600 430,clip]{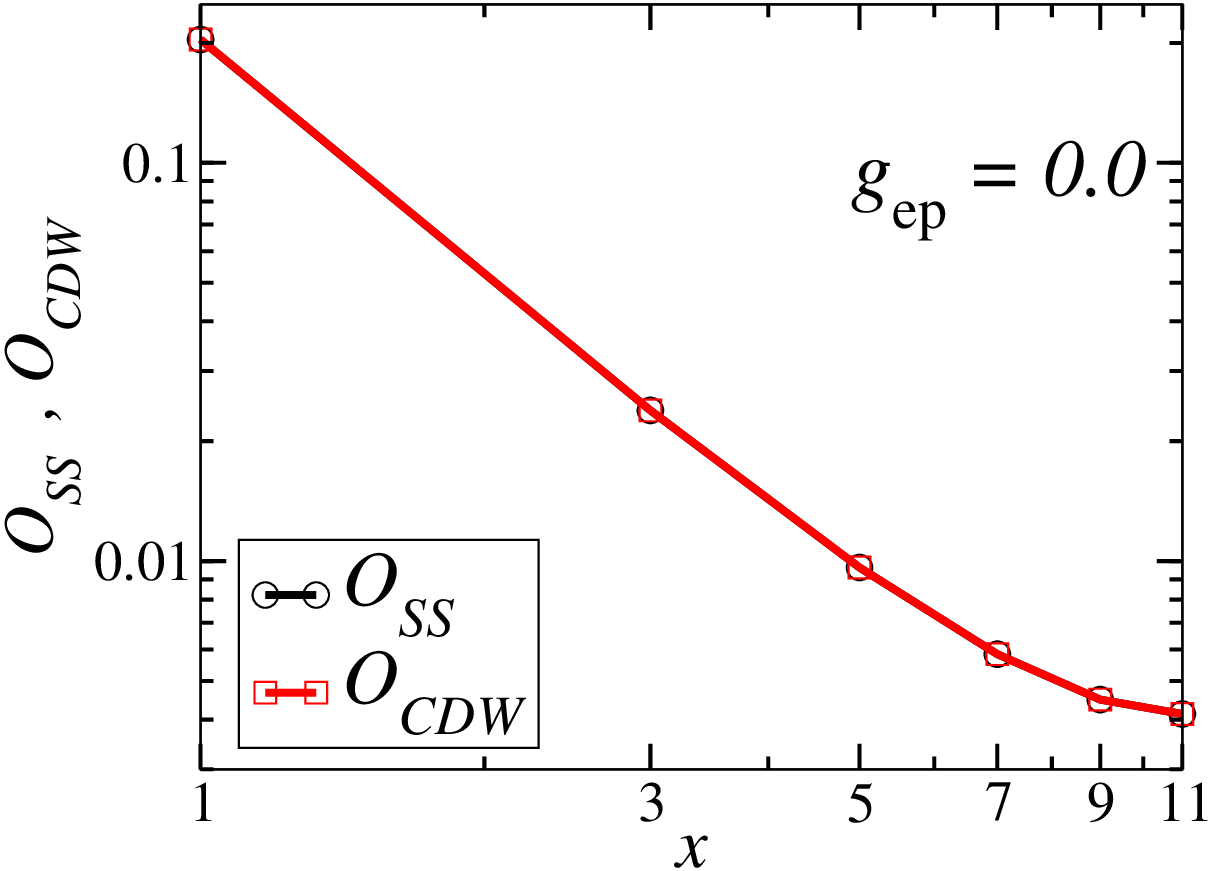} 
\includegraphics*[height=0.15\textheight,width=0.23\textwidth, viewport=-5 0 600 430,clip]{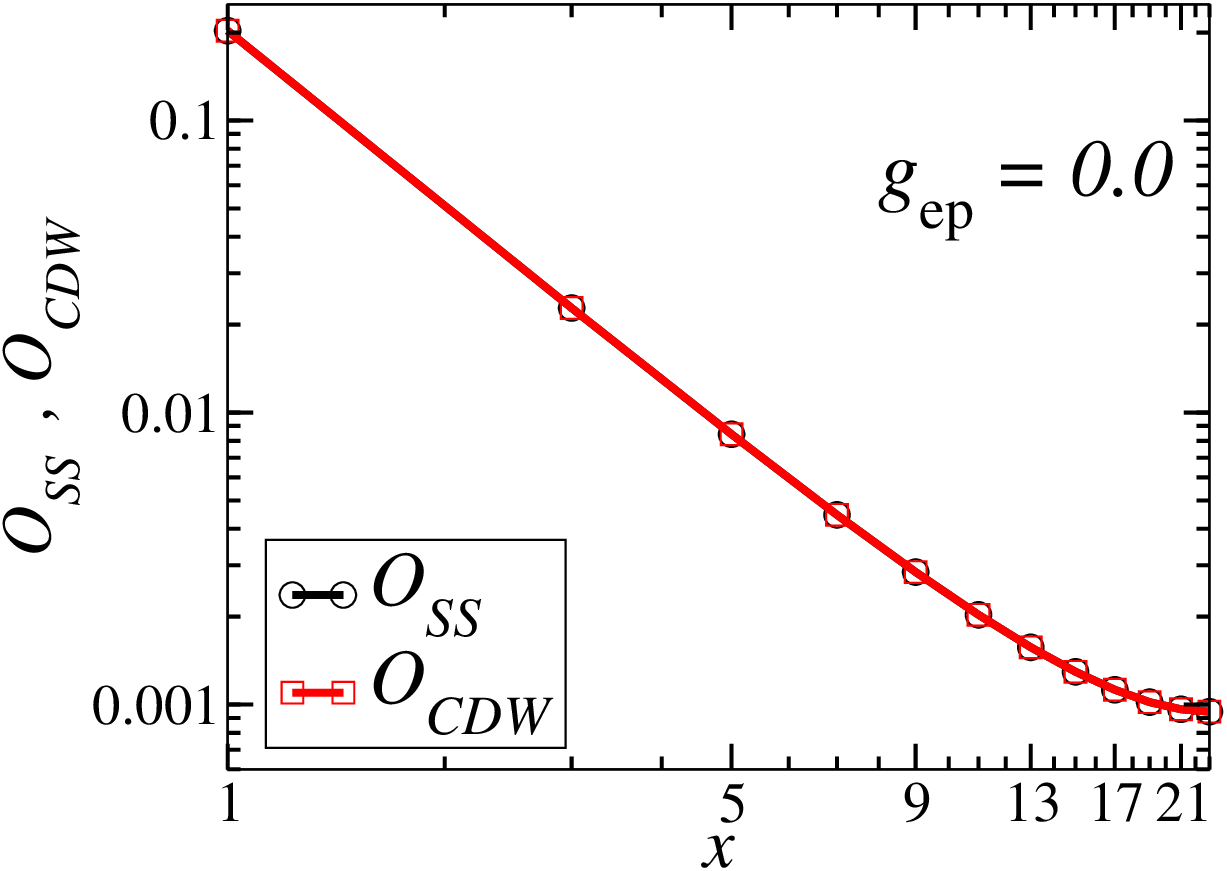}}
\centerline{
\includegraphics*[height=0.15\textheight,width=0.23\textwidth, viewport=-5 0 600 430,clip]{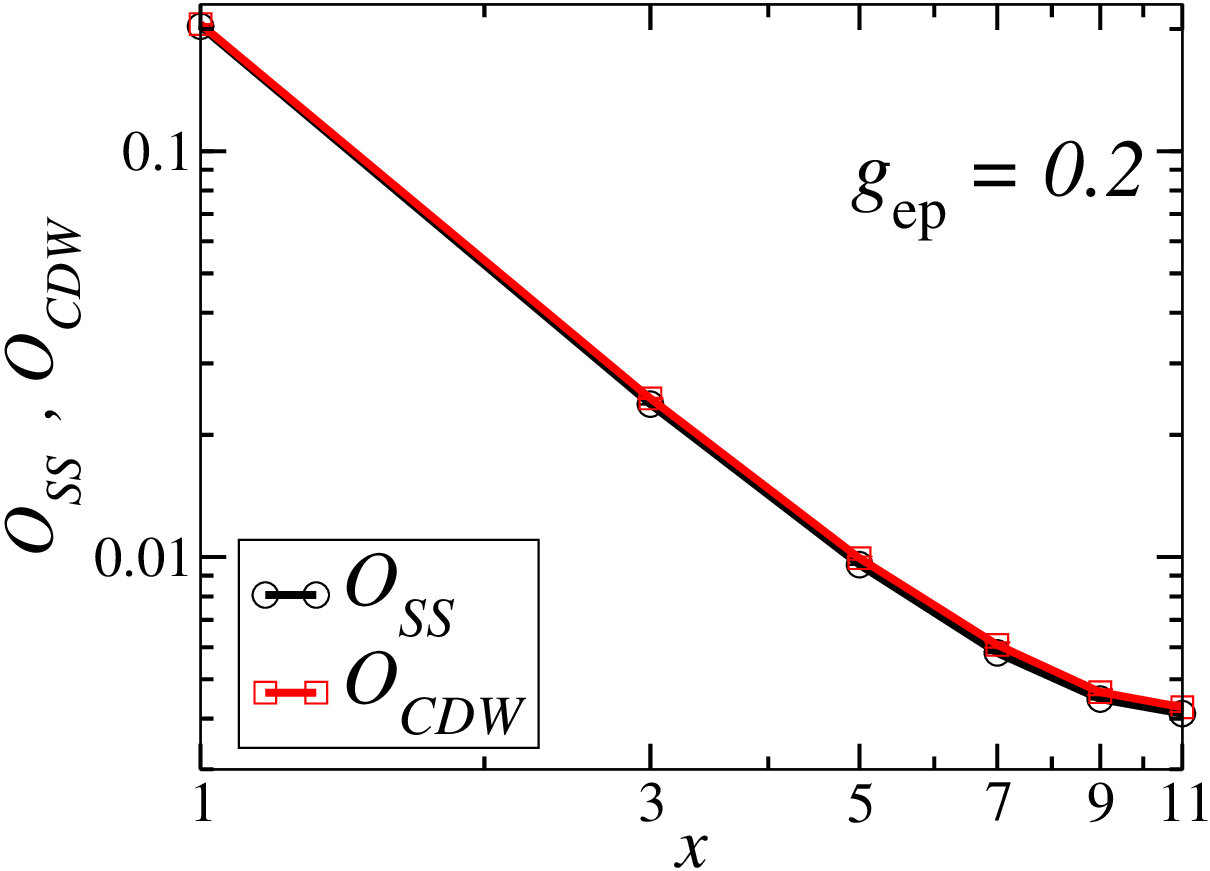} 
\includegraphics*[height=0.15\textheight,width=0.23\textwidth, viewport=-5 0 600 430,clip]{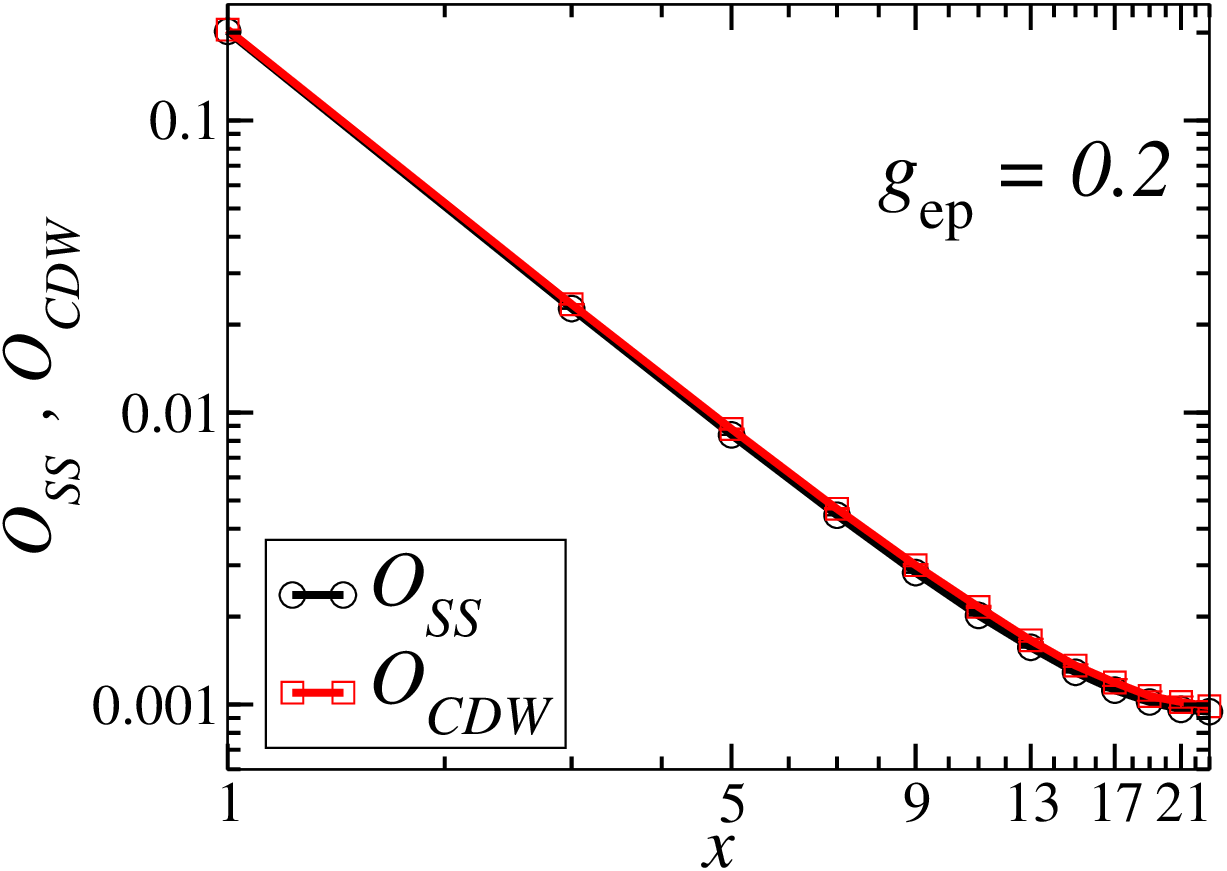}} 
\centerline{
\includegraphics*[height=0.15\textheight,width=0.23\textwidth, viewport=-5 0 600 430,clip]{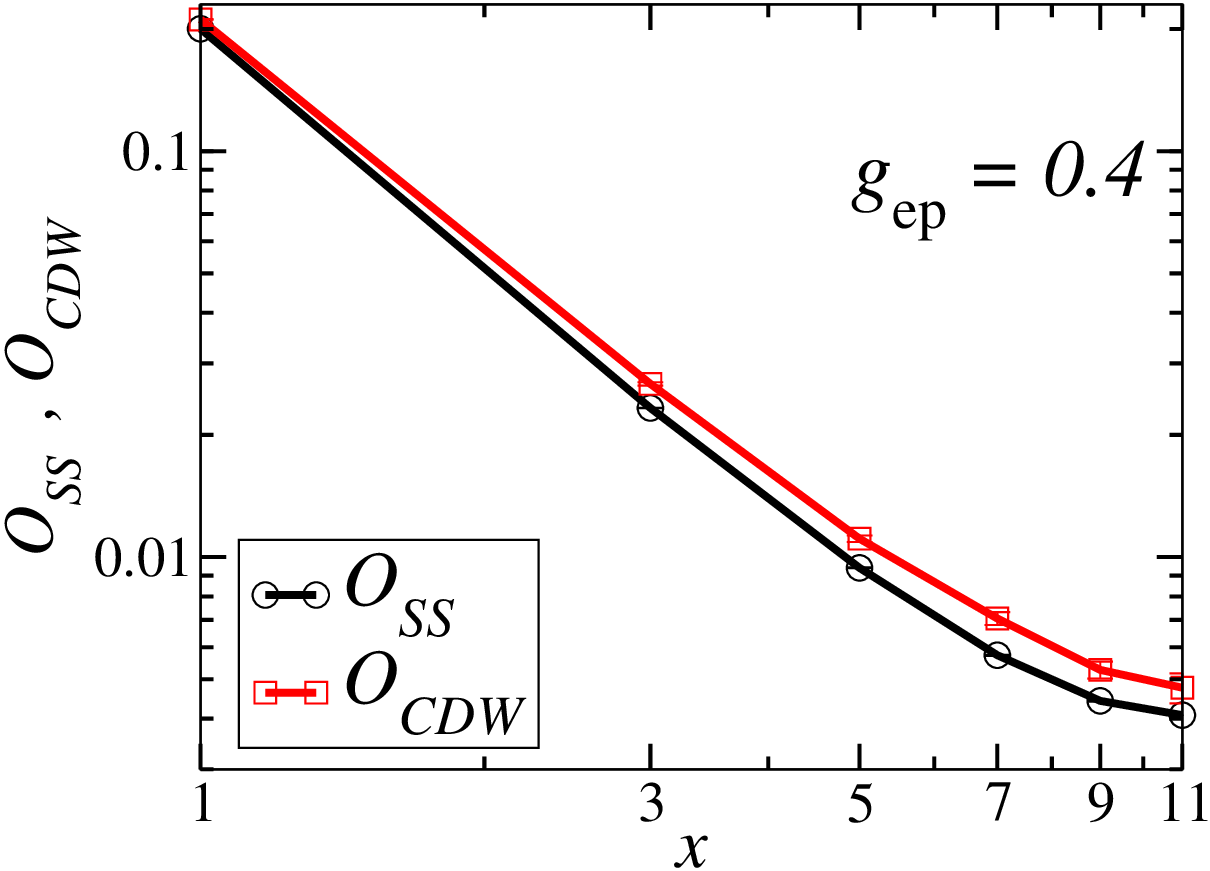} 
\includegraphics*[height=0.15\textheight,width=0.23\textwidth, viewport=-5 0 600 430,clip]{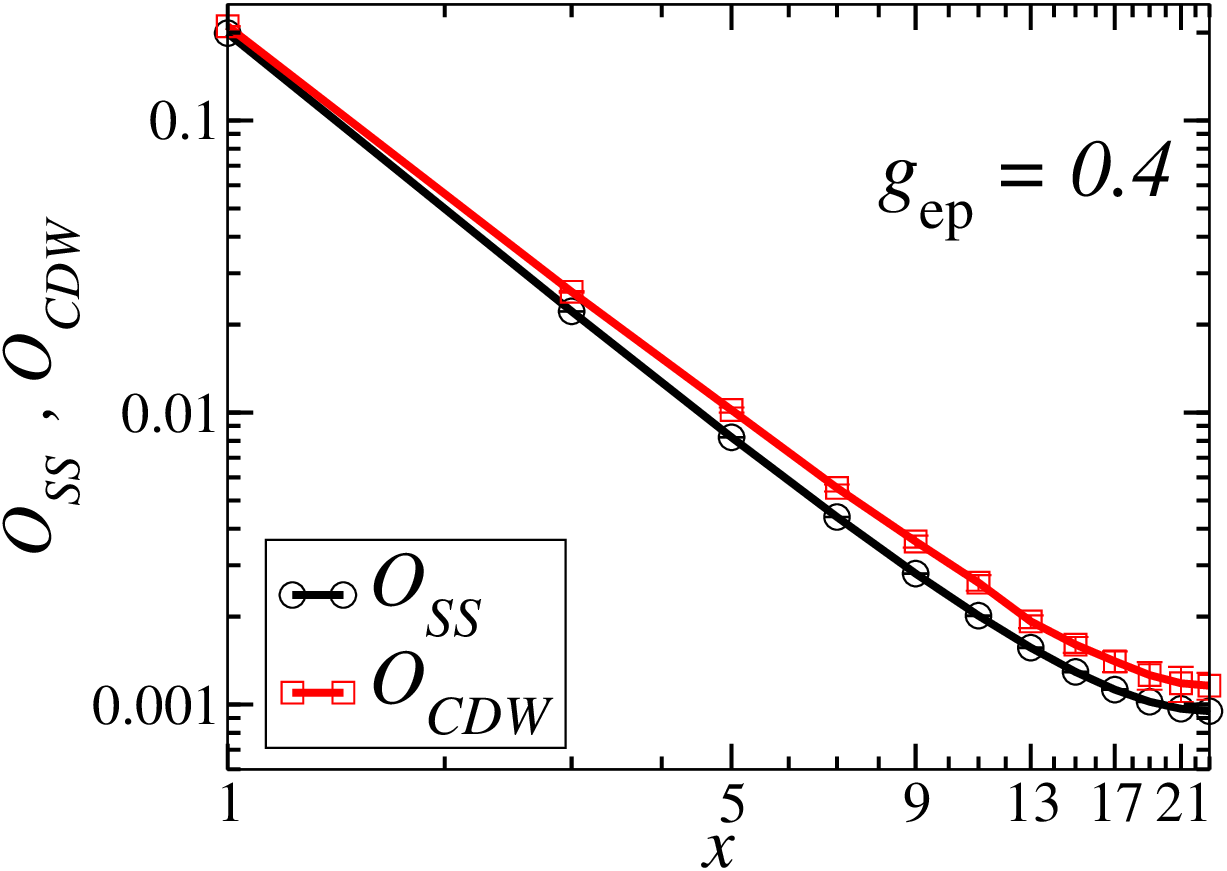}}
\centerline{
\includegraphics*[height=0.15\textheight,width=0.23\textwidth, viewport=-5 0 600 430,clip]{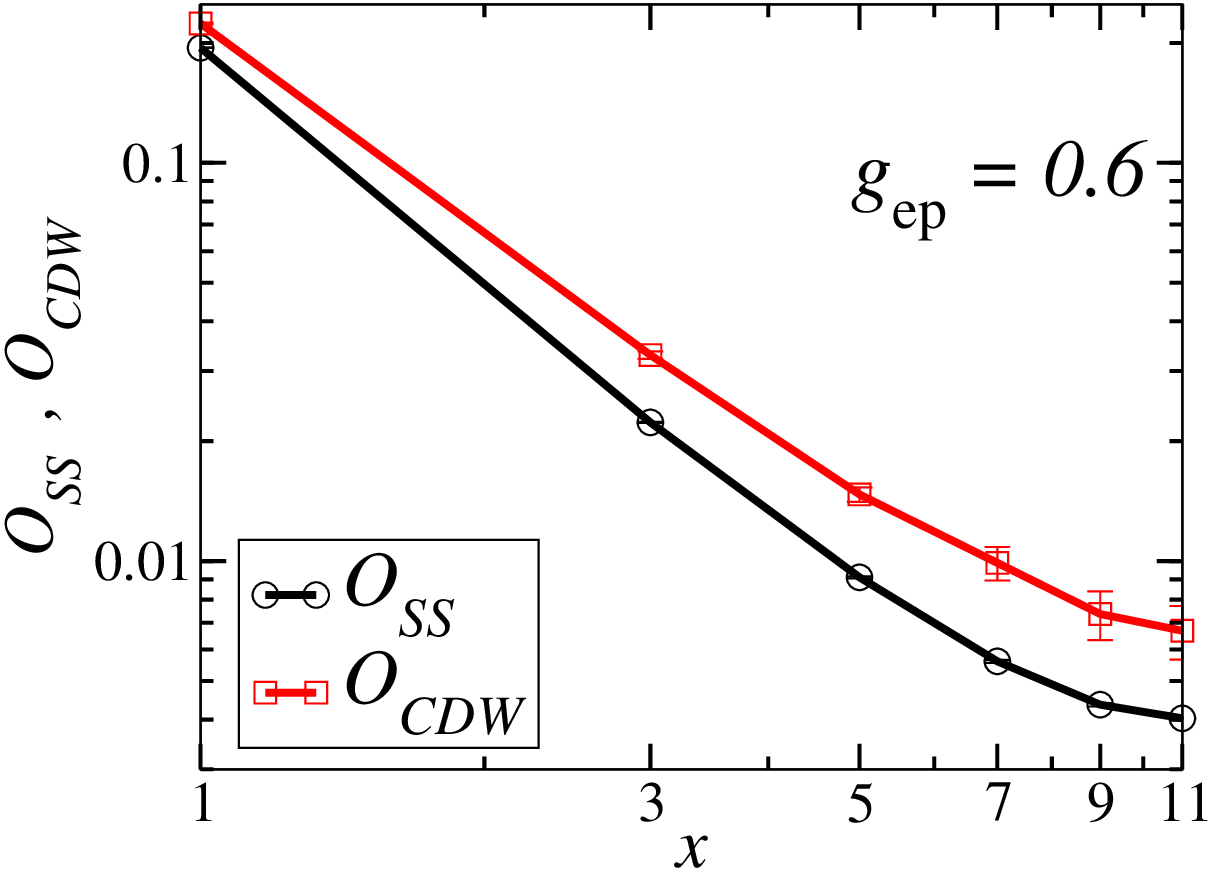} 
\includegraphics*[height=0.15\textheight,width=0.23\textwidth, viewport=-5 0 600 430,clip]{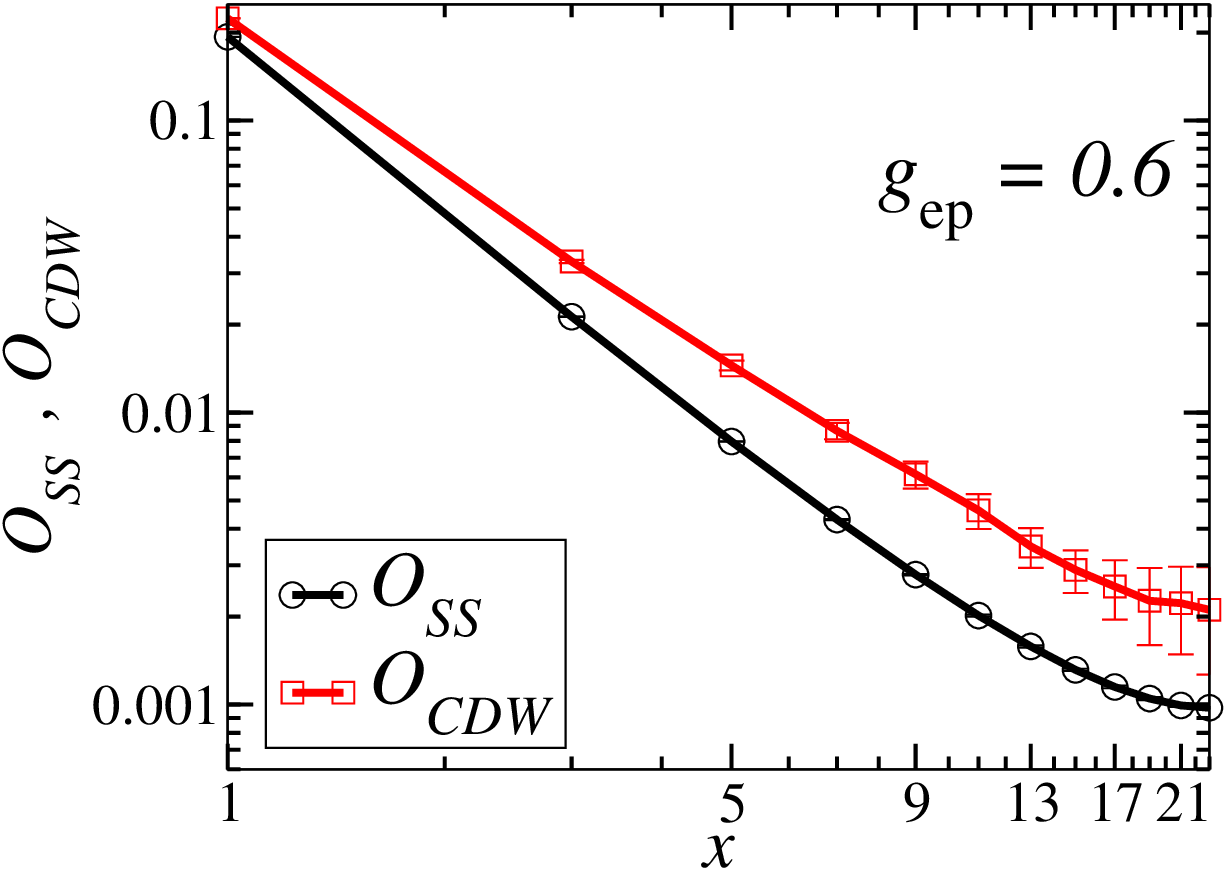}}
\centerline{
\includegraphics*[height=0.15\textheight,width=0.23\textwidth, viewport=-5 0 600 430,clip]{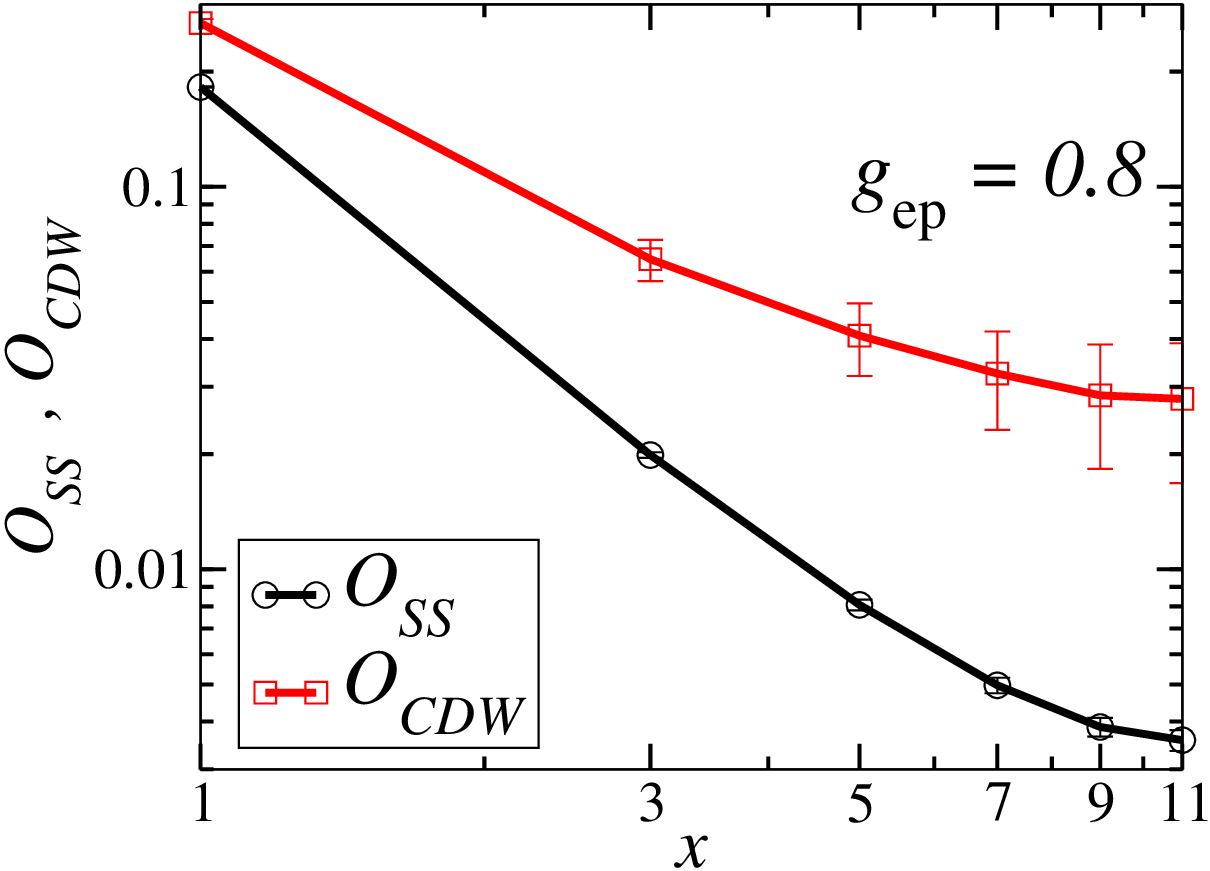} 
\includegraphics*[height=0.15\textheight,width=0.23\textwidth, viewport=-5 0 600 430,clip]{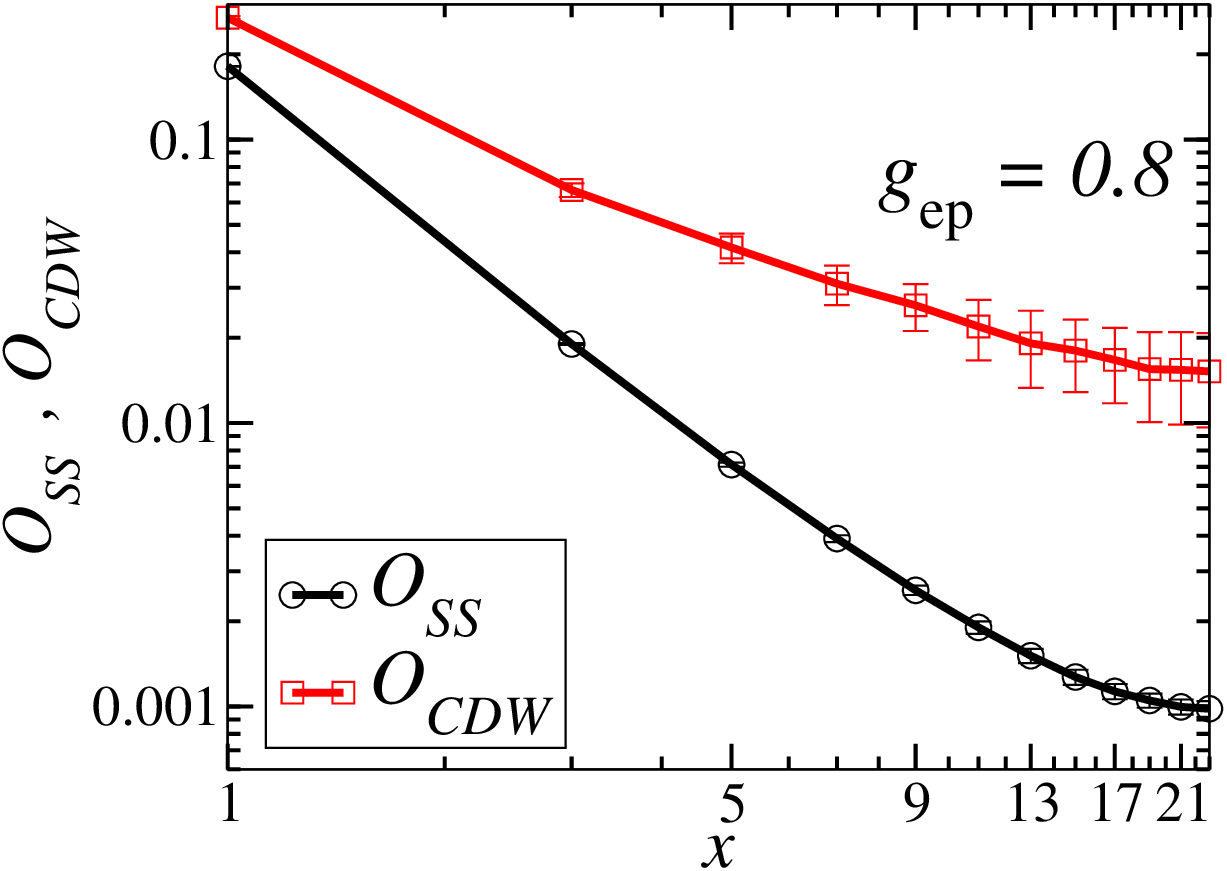}}
\centerline{
\includegraphics*[height=0.15\textheight,width=0.23\textwidth, viewport=-5 0 600 430,clip]{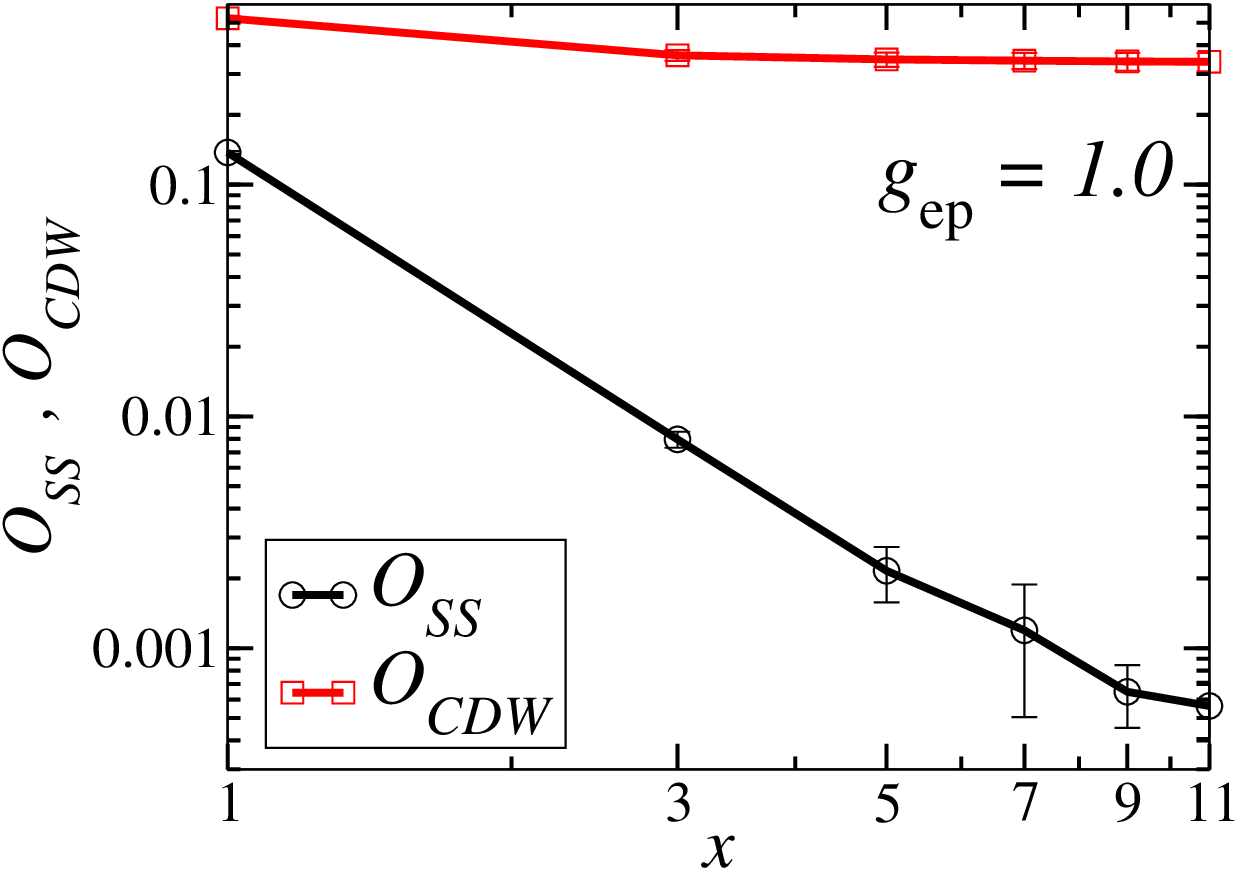} 
\includegraphics*[height=0.15\textheight,width=0.23\textwidth, viewport=-5 0 600 430,clip]{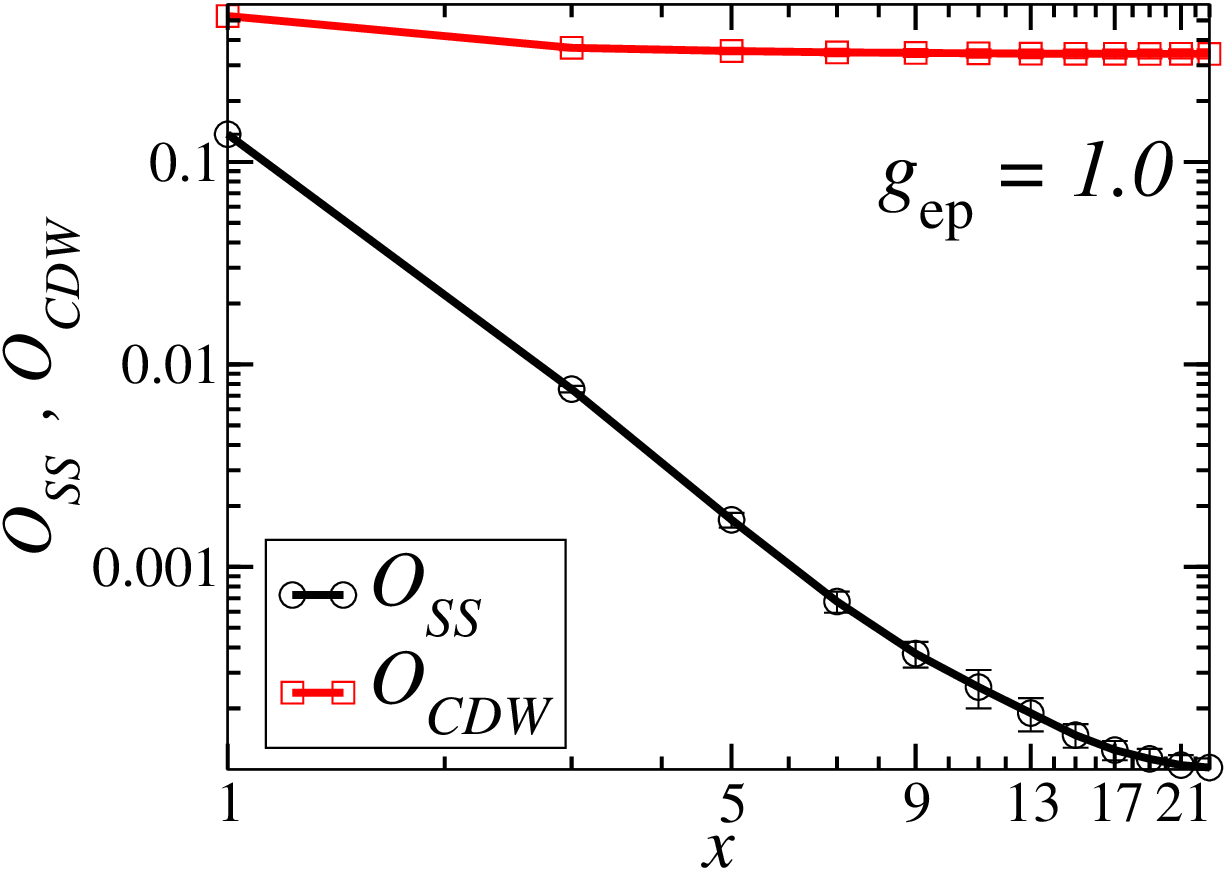}}

\caption{Singlet pairing correlations (open circles) and charge ordering correlations (open squares) functions as a function of real space distance for $g_{\rm{ep}}=0.0$ (the 1st row),
$g_{\rm{ep}}=0.2$ (the 2nd row), $g_{\rm{ep}}=0.4$ (the 3rd row), $g_{\rm{ep}}=0.6$ (the 4th row), $g_{\rm{ep}}=0.8$ (the 5th row) and $g_{\rm{ep}}=1.0$ (the 6th row). The left and right columns are for the lattices of $N = 22$ sites and $N = 46$ sites, respectively.}
\label{correlations_MC}
\end{figure}

\begin{eqnarray}
O^{CDW}(x)&=&\frac{1}{N}|\sum_{i=1}^{N} \langle n_{i} n_{i+x} - 1 \rangle| ;
\\
O^{SS}(x)&=&\frac{1}{N}|\sum_{i=1}^{N} \langle c_{i,\uparrow}^{\dagger} c_{i,\downarrow}^{\dagger} c_{i+x,\downarrow} c_{i+x,\uparrow} + H.c. \rangle| ,
\end{eqnarray}
where $N$ is the number of lattice sites. In the Monte Carlo calculation, we set $t=1$, $M=1/2$, and $\omega_{0}=1$, so that the $g_{\rm ep}$ defined in Eq. \ref{hhm} is the same as the
electron-phonon coupling $g$ defined in the Ref. [\onlinecite{Clay1}]. We show the correlation functions at different electron-phonon couplings in Fig. \ref{correlations_MC}.
At zero electron-phonon coupling, the Holstein model is a free fermion theory, and the singlet 
pairing and charge ordering are degenerate (the 1st row in the Fig. \ref{correlations_MC}). When the electron-phonon coupling is turned on, the charge ordering tends to decay more slowly, while the 
singlet pairing tends to decay faster, than in the free fermion case. The difference between them is expectedly small, possibly due to the exponentially small charge gap with respect to the electron-phonon 
coupling.\cite{Hirsch} Turning to larger electron-phonon couplings, the difference between the two different orderings becomes more obvious. As we are using systems with modest sizes (22 and 46 
sites with periodic boundary condition) only, it is difficult to determine whether the charge ordering is long range or decays as a power law, but the data clearly show that the charge ordering is 
enhanced and the singlet pairing is suppressed as the electron-phonon coupling is increased. For fairly large electron-phonon couplings, {\it e.g.}, $g_{\rm{ep}}=1.0$ (the 6th row in Fig. \ref{correlations_MC}), our data show a clearer signal that the charge ordering tends to become long range, as the charge correlation changes only very slowly beyond a few lattice sites. 

\begin{figure}[htb]
\centerline{
\includegraphics*[height=0.26\textheight,width=0.48\textwidth, viewport=0 0 620 450,clip]{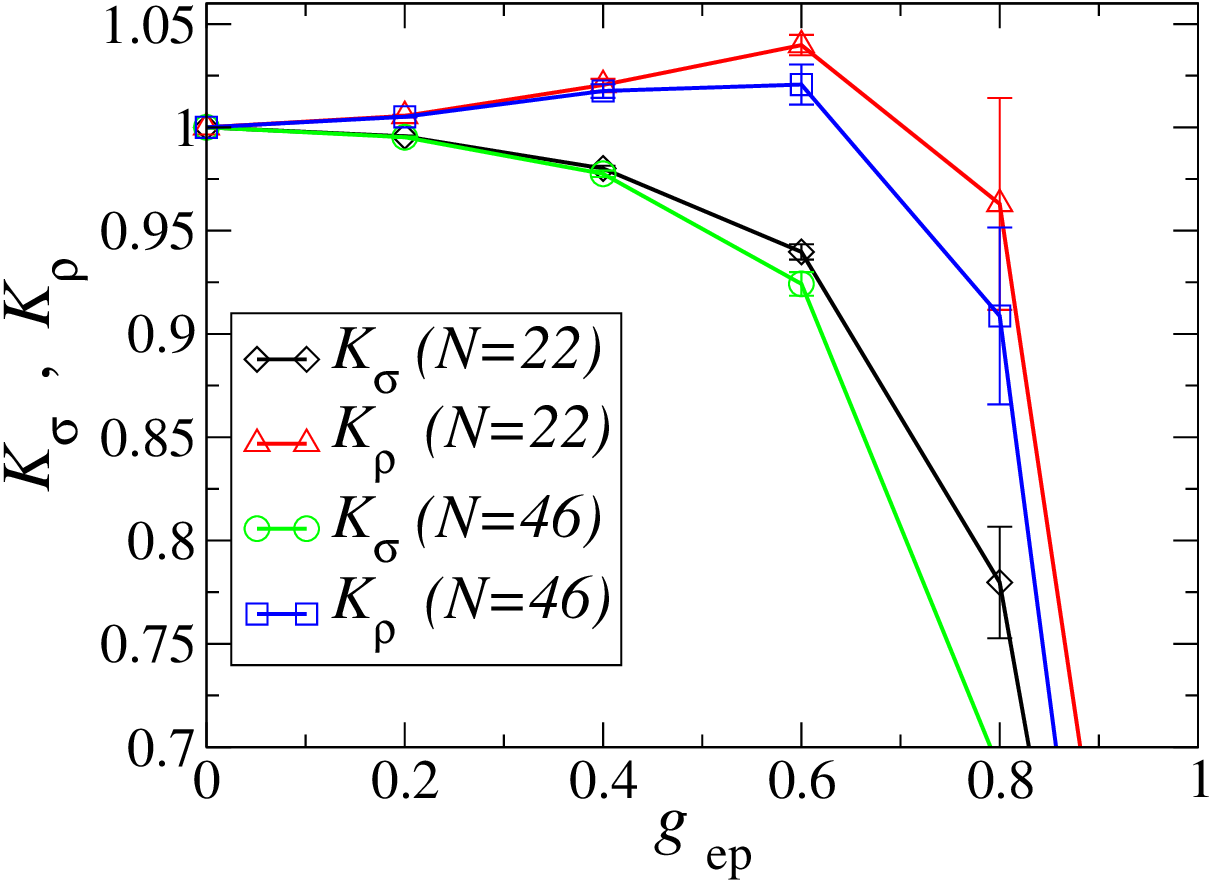}} 
\caption{The Luttinger charge exponents $K_{\rho}$ and spin exponents $K_{\sigma}$ for different electron-phonon couplings. $N$ is the number of lattice sites.}  
\label{K_MC}
\end{figure}

We also calculate the Luttinger charge exponent by the same method used previously for the Holstein model,\cite{Clay1} which found $K_{\rho} > 1$.  Using the set of data obtained in the same 
run for 
the correlation functions presented in Fig. \ref{correlations_MC}, we reproduce the result that the Luttinger charge exponent is clearly above {\it one} at least for $g_{\rm{ep}} \leq 0.6$ (see 
Fig. \ref{K_MC}). Using the LL relations would imply the dominant singlet pairing state for the model in this region; however, from the correlation functions we know that the singlet pairing decays 
faster than the charge ordering and is thus not dominant. The $K_{\rho} > 1$ obtained from {\it finite size} calculations is not incompatible with the CDW correlation decaying slower than that 
of the SS.

From our Monte Carlo data, we can conclude that singlet pairing is not the dominant instability for any range of electron-phonon coupling (although, strictly speaking, numerical data can 
hardly provide a conclusive answer for $g_{\rm ep} \rightarrow 0^{+}$). This provides a reasonably convincing answer to Question (II). The most prominent open question is 
whether the long range CDW ordering or equivalently the charge gap can be destroyed by the quantum fluctuations from the dynamical phonons as the electron-phonon coupling is decreased below a certain 
value. 

\section{Discussion}

Perturbative renormalization group approaches have consistently shown that the charge part is always 
gapped and thus the CDW is long range.\cite{Bourbonnais1,Tam,Bakrim,Bindloss,Bourbonnais2,Yonemitsu} This is ultimately related to the fact that the continuous spin rotation symmetry is broken by any 
finite 
phonon 
frequency which leads to a relevant Umklapp scattering. This result already appears in the perturbative renormalization group approaches within a second-order perturbation approximation.
For a metallic phase to be obtained in a perturbative renormalization group approach for the Holstein model, in particular for degenerate CDW and SS 
phase 
to be obtained, the summation of all the higher order terms in the perturbation series would have to
cancel exactly the second order contribution to restore the spin rotation symmetry. This is an unlikely scenario, and there is no convincing 
argument for why this should happen. 

On the other hand, to determine precisely the nature of the ground state from numerical calculations for very weak couplings is challenging, as the charge gap is exponentially small with 
respect to the electron-phonon coupling, even within the mean field approximation ($\Delta_{\rho} \sim exp(-1/g_{\rm ep}^{2})$). Quantum fluctuations from the dynamical phonons will further suppress 
it, especially at large phonon frequency.\cite{Hirsch,Schmeltzer,Bourbonnais1} Nonetheless, recent density matrix renormalization group studies did argue for the metallic 
phase, and further claimed that there are two types of metallic phases in both Holstein and Holstein-Hubbard model.\cite{Feshke2,Ejima} One of the metallic phases is gapless in both the spin and 
charge sectors, which normally implies that either triplet superconducting pairing or spin density wave ordering is the dominant power-law decaying ordering\cite{Voit}, even though spin fluctuations are 
completely absent in the Holstein model. More elaborate calculations using Monte Carlo algorithms which can directly extract the pairing correlations for much larger 
system sizes may provide further insight.\cite{Rousseau}

Although we cannot rule out the possibility of a metallic phase from this study, it is quite clear that the Holstein model in one dimension does not support dominant superconducting 
correlations for the half-filled case. This is due to the relevant Umklapp scattering from the perspective of the perturbative renormalization group,\cite{Bourbonnais1,Caron,Bakrim,Tam} which is 
supposed to be reliable from weak to intermediate coupling. In general, breaking the particle-hole symmetry alone,\cite{Tezuka3} 
by, for example, including non-linear term for the phonon, is probably not 
sufficient to induce dominant superconducting correlation.\cite{Freericks} The crucial point is to weaken the contribution from the particle-hole channel by removing the nesting 
condition. Theoretically, this can be achieved by introducing long range hopping terms or doping away from the 
half-filling.\cite{Tezuka3,Clay2,Tam2}

\section{Conclusion}

In this paper we have examined the nature and applicability of using the LL relations to determine the possible existence and the nature of a proposed metallic ground state of the one-dimensional half-filled Holstein model. We conclude by reiterating our answers to the three questions 
posed in our introduction. (I) The LL scaling relation $\alpha_{CDW} \alpha_{SS} = 1$ is valid only in the limit of infinite size systems; for finite size systems there are corrections that 
lead to $K_{\rho} > 1$.  (II) For a wide range of electron-phonon coupling, the dominant correlation for finite size systems in the Holstein model is the CDW. (III) Although the CDW seems to be always the dominant correlation, without a direct 
study of the gap--which is expected to be exponentially small with respect to the electron-phonon coupling and thus very challenging to study numerically--we cannot rule out the possibility of a 
charge gapless ``metallic" phase for weak electron-phonon coupling.

\section{Acknowledgments}
We thank Antonio Castro Neto, R. Torsten Clay and Anders Sandvik for valuable discussions,  the Aspen Center for Physics where some of the work was accomplished, and the Center for Computational Science at Boston University for its partial support of the computational work. S-W Tsai gratefully acknowledges support from NSF under grant DMR0847801 and from the UC-Lab FRP under award number 09-LR-05-118602.



\begin{thebibliography}{99}

\bibitem{Peierls} R. Peierls, Quantum Theory of Solids (Oxford: Pergamon, 1955).

\bibitem{Holstein1} T. Holstein, Ann. Phys. {\bf 8}, 325 (1959).

\bibitem{Holstein2} T. Holstein, Ann. Phys. {\bf 8}, 343 (1959).

\bibitem{Hirsch} J. E. Hirsch and E. Fradkin, Phys. Rev. B {\bf 27}, 4302 (1983).
  
\bibitem{Hirsch2} J. E. Hirsch, Phys. Rev. B {\bf 31}, 6022 (1985).  

\bibitem{Caron} L. G. Caron and C. Bourbonnais, Phys. Rev. B {\bf 29}, 4230 (1984).

\bibitem{Bourbonnais1} C. Bourbonnais and L. G. Caron, J. Phys. France {\bf 50}, 2751 (1989). 

\bibitem{Schmeltzer} D. Schmeltzer, J. Phys. C: Solid State Phys. {\bf 20}, 3131 (1987).

\bibitem{Yonemitsu} K. Yonemitsu and M. Imada, Phys. Rev. B {\bf 54}, 2410 (1996). 

\bibitem{Wu} C. Q. Wu, Q. F. Huang, and X. Sun, Phys. Rev. B {\bf 52}, R15683 (1995).

\bibitem{Takada2} Y. Takada, J. Phys. Soc. Jpn. {\bf 65}, 1544 (1996).

\bibitem{Hotta} T. Hotta and Y. Takada, Physica B {\bf 230}, 1037 (1997).

\bibitem{Jeckelmann} E. Jeckelmann, C. Zhang, and S. R. White, Phys. Rev. B {\bf
  60}, 7950 (1999). 

\bibitem{Takada} Y. Takada and A. Chatterjee, Phys. Rev. B {\bf 67}, 081102(R)
  (2003).

\bibitem{Feshke3} H. Fehske, A.P. Kampf, M. Sekania, and G. Wellein, Eur. Phys. J. B {\bf 31}, 11 (2003).

\bibitem{Feshke} H. Fehske, G. Wellein, G. Hager, A. Wei\ss e, and A. R. Bishop, Phys Rev. B {\bf 69}, 165115 (2004).

\bibitem{H-Zhao} H. Zhao, C. Q. Wu, and H. Q. Lin, Phys. Rev. B {\bf 71}, 115201 (2005).

\bibitem{Tezuka} M. Tezuka, R. Arita, and H. Aoki, Physica B {\bf 359}, 708 (2005).

\bibitem{Tezuka2} M. Tezuka, R. Arita, and H. Aoki, 
  Phys. Rev. Lett. {\bf 95}, 226401 (2005). 

\bibitem{Tezuka3} M. Tezuka, R. Arita, and H. Aoki, Phys. Rev. B {\bf 76}, 155114 (2007).   

\bibitem{Feshke2} H. Fehske, G. Hager, and J. Jeckelmann, Europhys. Lett. {\bf 84}, 57001 (2008). 

\bibitem{Ejima} S. Ejima and H. Fehske, J. Phys.: Conf. Ser. {\bf 200}, 012031 (2010).

\bibitem{Chatterjee} A. Chatterjee, Adv. Condens. Matter Phys. {\bf 2010}, 350787 (2010). 

\bibitem{Zhao} J. Zhao and K. Ueda, J. Phys. Soc. Jpn. {\bf 79}, 074602 (2010).

\bibitem{Clay1} R. T. Clay and R. P. Hardikar, Phys. Rev. Lett. {\bf 95}, 096401 (2005).

\bibitem{Schulz} H. J. Schulz. Phys. Rev. Lett. {\bf 64}, 2831 (1990).

\bibitem{Yang} C. N. Yang, Rev. Mod. Phys. {\bf 34}, 694 (1962).

\bibitem{Anderson1} P. W. Anderson, Phys. Rev. {\bf 110}, 827 (1958).

\bibitem{Anderson2} P. W. Anderson, Phys. Rev. {\bf 130}, 439 (1963).

\bibitem{Mermin-Wagner} N. D. Mermin and H. Wagner, Phys. Rev. Lett. {\bf 17}, 1133 (1966).

\bibitem{Hohenberg} P. C. Hohenberg, Phys. Rev. {\bf 158}, 383 (1967).

\bibitem{Coleman} S. Coleman, Comm. Math. Phys. {\bf 31}, 259 (1973).
  
\bibitem{Clay2} R. P. Hardikar and R. T. Clay, Phys. Rev. B {\bf 75}, 245103 (2007).

\bibitem{Guinea} F. Guinea, J. Phys. C: Solid State Phys. {\bf 16}, 4405 (1983). 

\bibitem{Loss} D. Loss and T. Martin, Phys. Rev. B {\bf 50}, 12160 (1994).

\bibitem{Martin}T. Martin and D. Loss, Int. J. Mod. Phys. B {\bf 9}, 495 (1995).
   
\bibitem{Tam} K.-M. Tam, S.-W. Tsai, D. K. Campbell, and A. H. Castro Neto, Phys. Rev. B {\bf 75}, 161103(R) (2007).

\bibitem{Solyom}J. S\'{o}lyom, Adv. Phys. {\bf 28}, 201 (1979).
  
\bibitem{Voit} J. Voit, Rep. Prog. Phys. {\bf 58}, 977 (1995).

\bibitem{Emery} V. J. Emery, in {\it Highly Conducting One-Dimensional
    Solids}, p. 327, edited by J. T. Devreese, R. P. Evrard, and V. E. van Doren
   (Plenum, New York, 1979).

\bibitem{Luther-Emery} A. Luther and V. J. Emery, Phys. Rev. Lett. {\bf 33}, 589 (1974).

\bibitem{Chen} Y. Chen, D. K. K. Lee, and M. U. Luchini, Phys. Rev. B {\bf 38}, 8497 (1988).

\bibitem{Grest} G. S. Grest, E. Abrahams, S.-T. Chui, P. A. Lee, and A. Zawadowski, Phys. Rev. B {\bf 14}, 1225 (1976).

\bibitem{Bourbonnais2} C. Bourbonnais and L. G. Caron, Int. J. Mod. Phys. B {\bf 5}, 1033 (1991). 

\bibitem{Zimanyi} G. T. Zimanyi, S. A. Kivelson, and A. Luther, Phys. Rev. Lett. {\bf 60}, 2089 (1998).

\bibitem{Bindloss} I. P. Bindloss, Phys. Rev. B {\bf 71}, 205113 (2005).

\bibitem{Tsai} S.-W. Tsai, A. H. Castro Neto, R. Shankar, and D. K. Campbell, Phys. Rev. B {\bf 72}, 054531 (2005).
  
\bibitem{Tsai2} S.-W. Tsai, A. H. Castro Neto, R. Shankar, and D. K. Campbell, Phil. Mag. {\bf 86}, 2631 (2006).

\bibitem{Tsai3} S.-W. Tsai, A. H. Castro Neto, R. Shankar, and D. K. Campbell, J. Phys. Chem. Solids {\bf 67}, 516 (2006).
  
\bibitem{Bakrim} H. Bakrim and C. Bourbonnais, Phys. Rev. B {\bf 76}, 195115 (2007). 

\bibitem{Blankenbecler} R. Blankenbecler, D. J. Scalapino, and R. L. Sugar, Phys. Rev. D {\bf 24}, 2278 (1981). 

\bibitem{White} E. Y. Loh, Jr., J. E. Gubernatis, R. T. Scalettar, S. R. White, D. J. Scalapino, and R. L. Sugar, Phys. Rev. B {\bf 41}, 9301 (1990).

\bibitem{Berger} E. Berger, P. Val\'a\v{s}ek, and W. von der Linden, Phys. Rev. B {\bf 52}, 4806 (1995).

\bibitem{Linden} W. von der Linden, E. Berger, and P. Val\'a\v{s}ek, J. Low Temp. Phys. {\bf 99}, 517 (1995). 

\bibitem{Rousseau} V. G. Rousseau, Phys. Rev. E {\bf 78}, 056707 (2008). 

\bibitem{Freericks} J. K. Freericks and G. D. Mahan, Phys. Rev. B {\bf 54}, 9372 (1996). 

\bibitem{Tam2} Ka-Ming Tam, S.-W. Tsai, D. K. Campbell, and A. H. Castro Neto, Phys. Rev. B {\bf 75}, 195119 (2007).


\end{thebibliography}
\end{document}